\documentclass[12pt]{article}
\usepackage{graphicx}
\usepackage{amssymb}
\def\lsim{\mathrel{\raise.3ex\hbox{$<$\kern-.75em\lower1ex\hbox{$\sim$}}}}
\def\gsim{\mathrel{\raise.3ex\hbox{$>$\kern-.75em\lower1ex\hbox{$\sim$}}}}

\newcommand{\neb}{{\bar{\nu}_e}}

\def\vev#1{{\langle#1\rangle}}

\begin{document}

\title{Supernova neutrinos can tell us the neutrino mass hierarchy independently of flux models}
\author{V. Barger$^1$, Patrick Huber$^1$ and Danny Marfatia$^{2}$\\[2ex]
\small\it $^1$Department of Physics, University of Wisconsin, Madison, WI 53706\\
\small\it $^2$Department of Physics and Astronomy, University of Kansas, Lawrence, KS 66045}

\date{}

\maketitle

\begin{abstract}
We demonstrate that the detection of shock modulations of the 
neutrino spectra from a galactic
core-collapse supernova is sufficient to obtain a high 
significance determination of the neutrino mass hierarchy if the supernova
event is observed in both a Mton-class water Cherenkov detector and 
a 100 kton-class liquid argon detector. 
Neither detailed supernova neutrino flux modelling nor 
observation of Earth matter effects is needed for this determination. 
As a corollary, a nonzero value of $\theta_x$ will be established.

\end{abstract}

\newpage

\section{Introduction}

The current status of neutrino oscillation parameter estimations can be 
very briefly summarized~\cite{Barger:2003qi} 
as follows: Atmospheric (solar) neutrinos oscillate with 
$|\delta m^2_a|\sim 0.002$ eV$^2$  
and $\theta_a \sim \pi/4$~\cite{superkatm} 
($\delta m^2_s \sim 8\times 10^{-5}$ eV$^2$, 
$\theta_s \sim \pi/6$~\cite{Eguchi:2002dm}){\footnote{In our
notation, $\delta m^2_s$ ($\delta m^2_a$) is the solar (atmospheric) 
mass-squared difference and $\theta_s$,
$\theta_x$ and $\theta_a$ are the mixing angles conventionally denoted 
by $\theta_{12}$, $\theta_{13}$ and $\theta_{23}$, 
respectively~\cite{Barger:2003qi}.}}. All we presently know
about $\theta_x$ is that
$\sin^2 \theta_x \lsim 0.05$ at the 2$\sigma$ C.~L.~\cite{Apollonio:1999ae}.
A long-standing hope is that neutrinos from a core-collapse supernova (SN) may shed
light on two of the unknown oscillation parameters, sgn($\delta m^2_a$) and 
$\theta_x$. 

Only a handful of neutrinos from a Type II SN have ever been 
detected. 
The detection of 11 neutrinos from SN 1987A
in Kamiokande II~\cite{k2sn} and 8 neutrinos in the Irvine Michigan
Brookhaven experiment~\cite{imbsn} 
have been of great importance for understanding
core-collapse~\cite{loredo}. It is evident that
the physics potential offered by a future galactic SN event
is immense. With cognizance of this potential, experiments dedicated
to SN neutrino detection have been proposed~\cite{uno} even though
only a few galactic SN are expected per century.

Attempts have been made to extract neutrino oscillation parameters
from the 19 SN 1987A events.  However, conclusions drawn from such 
analyses are highly dependent on the neutrino flux model 
adopted and are far from robust.
For example (and within the context of this paper), 
it was claimed that the data favor the normal hierarchy ($\delta m^2_a>0$)
over the inverted hierarchy ($\delta m^2_a<0$) provided $\sin^2 \theta_{x} \gsim
10^{-4}$~\cite{minnun}, but this conclusion was
contradicted in Ref.~\cite{us}.

Neutrinos from a galactic SN could in principle provide a wealth of 
information on neutrino
oscillations. A determination of $\theta_{x}$ and the neutrino mass
hierarchy from SN neutrinos is unique in that 
ambiguities~\cite{ambiguities,bmw} arising
from the unknown $CP$ phase
$\delta$ and the deviation of atmospheric neutrino mixing from maximality
 do not corrupt it.
The absence of the eight-fold parameter degeneracies that are
inherent in long baseline experiments~\cite{bmw} results because (i)
nonelectron neutrino fluxes{\footnote{We focus on detection via 
charged current $\nu_e$ and $\bar{\nu}_e$ interactions,
which cannot distinguish between the
different nonelectron neutrino species (that we denote by
 $\nu_x$ with $x=\mu,\tau, \bar{\mu}, \bar{\tau}$).}} 
do not depend on $\delta$
independently of neutrino conversion~\cite{newmodel4},
and so SN neutrinos directly probe $\theta_{x}$, and (ii)
whether atmospheric mixing is maximal or not is immaterial since
$\theta_a$ does not affect the oscillation dynamics.

Investigations of the effect of neutrino oscillations on SN neutrinos in the 
context of a static density profile ({\it {i.e.,}} neglecting shock effects) 
have been made in Refs.~\cite{dighe,futuresn,us2}.
Whether or not the mass hierarchy can
be determined and $\theta_x$ be constrained depends sensitively on the 
strength of
the hierarchy between $\vev{E_{\bar{\nu}_e}}$ and $\vev{E_{\nu_x}}$.
 The higher  
$\vev{E_{\nu_x}}/\vev{E_{\bar{\nu}_e}}$ is above unity, the
better the possible determinations~\cite{dighe}. Unfortunately, modern SN models
that include all relevant neutrino interaction effects like nuclear recoil
and nucleon bremsstrahlung indicate that the hierarchy of average energies is
likely smaller than expected from traditional predictions; 
$\vev{E_{\nu_x}}/\vev{E_{\bar{\nu}_e}}$ is expected to be about 1.1, 
and no larger than 1.2~\cite{raffsn} as opposed to ratios above 1.5 from
older SN codes~\cite{models1}. 
Another relevant uncertainty is that different
SN models predict different degrees to which
 equipartitioning of energy between $\nu_e$, $\bar{\nu}_e$ and $\nu_x$ 
is violated. 
For example, in Ref.~\cite{totani} an almost perfect equipartitioning
is obtained while according to Refs.~\cite{newmodel5,mezzacappa},
equipartitioning holds only to within a factor of 2.

Given these uncertainties, it is not a simple task to determine $\theta_x$
and the mass hierarchy simultaneously from SN data~\cite{us2}. 
A significant improvement would be a 
determination of the mass hierarchy independently of predictions for
$\vev{E_{\nu_x}}/\vev{E_{\bar{\nu}_e}}$ and equipartitioning from 
SN models. 
In this paper we propose a new method using SN neutrinos to determine the
mass hierarchy that exploits recent advances
in the understanding of shock propagation in SN. 

At densities $\sim 10^3$ g/cm$^3$, neutrino oscillations are governed by 
$\delta m^2_{a}$ and $\sin^2 \theta_{x}$~\cite{dighe}. 
Neutrinos (antineutrinos) pass 
through a resonance if  $\delta m^2_{a}>0$ ($\delta m^2_{a}<0$). 
As the shock traverses the 
resonance, adiabaticity is severely affected causing oscillations
 to be temporarily suppressed, as first pointed out in Ref.~\cite{fuller}. After the shock moves beyond the resonance,
oscillations are restored. Then one expects a dip in the time evolution of 
the average neutrino energy and the number of 
events{\footnote{Recently, 
this idea was taken one step further in Ref.~\cite{tomas}.
A reverse shock caused by the collision between a neutrino-driven baryonic wind
and the more slowly moving primary ejecta may also form. 
The direct and reverse shocks yield a ``double dip'' 
signature~\cite{tomas}. 
In the present work we restrict our attention to the effects of the 
forward shock which is a generic feature of SN models and
whose existence is better established than that of the reverse 
shock.}}.
This modulation is visible in the neutrino (antineutrino) channel for a
normal (inverted) mass hierarchy and only if 
$\tan^2 \theta_{x} \gg 10^{-5}$ {\it {i.e.}}, only for oscillations that would
occur adiabatically for a static density profile. Previous work exploiting
the dip to obtain information about oscillation parameters can be found in 
Ref.~\cite{catchall}.

Within the first 4 seconds or so, the violation of adiabaticity caused by the 
shock is felt only by neutrinos with energy less than about 20 MeV. 
At these energies some models predict the $\nu_x$ flux
to be larger than the $\nu_e$ and $\bar{\nu}_e$ fluxes~\cite{newmodel5} 
and others predict
the converse~\cite{totani}. At later times, the shock
affects higher energy neutrinos for which all models predict the $\nu_x$ flux
to be dominant.
The dip is observable even for
$\vev{E_{\nu_x}}=\vev{E_{\bar{\nu}_e}}$ because the fluxes are 
flavor-dependent~\cite{tomas}. Thus, a signature in high energy neutrinos a few
seconds after bounce is quite model-independent. 
Through our analysis we
show that the signal
is so robust that a restriction to high-energy events is unnecessary.

We investigate the significance with which the mass hierarchy can 
be determined by measurement of 
the $\nu_e$ spectrum at a 100 kton liquid argon 
detector and  the $\bar{\nu}_e$ spectrum at a 1 Mton water 
Cherenkov detector from a galactic SN at a distance 
of 10 kpc with binding energy $3\times 10^{53}$ ergs. The number of 
unoscillated events in the liquid Ar (water Cherenkov) detector is expected
to be ${\cal{O}}(10^5)$ (${\cal{O}}(10^6)$). 
(Although a liquid argon detector can measure both the $\nu_e$ and 
$\bar{\nu}_e$ spectra,  
it is an order of magnitude more sensitive to the $\nu_e$ 
flux than to the $\bar{\nu}_e$ flux~\cite{bueno}). We are interested 
in the detectability of a dip in the time evolution of 
either the $\nu_e$ or the $\bar{\nu}_e$ spectrum, but not both.
By correlating the two spectra, it should be possible to establish the mass 
hierarchy and that $\tan^2 \theta_{x} \gg 10^{-5}$. The presence of a 
dip will also provide further evidence that SN simulations correctly depict 
shock propagation.
 If a dip is not found in either channel, it would  
suggest that $\tan^2 \theta_{x} \lsim 10^{-5}$, since the salient features
of the theory of core-collapse SN have already received strong support 
from SN 1987A~\cite{loredo}.

\section{Shock density profile and neutrino oscillation probabilities} 

Realistic time-dependent density profiles of SN are obtained from detailed
numerical siumulations. As one moves in towards the neutron star, 
the profile at a given instant has a sharp
density rise followed by a rarefaction region where the density can drop
significantly below that at the outer edge of the shock.
The authors of Ref.~\cite{fuller} have provided a generic time-dependent 
density profile that mimics those of supernova simulations. 
The shock front is steepened artificially to reintroduce the physical 
requirement of a density discontinuity 
which is lost in hydrodynamic simulations
due to the limited (few 100 km) spatial resolution. 
We adopt the empirical parameterization of this profile 
(which is continuous in the supernova radius and time) 
developed in Ref.~\cite{fogli}. 

Because of the dip in density in the rarefaction region, neutrinos may hop
between mass eigenstates up to 3 times before leaving the SN envelope. 
Under the assumptions that the
transitions factorize and the neutrino phases can be averaged away, 
a simple analytic expression for the overall
hopping probability $P_H(|\delta m^2_a|,\theta_x)$ 
was obtained in Ref.~\cite{fogli}, 
which agrees remarkably well with phase-averaged
results of Runge-Kutta evolution of the neutrino flavor propagation equations.
We employ the analytic expression for $P_H$ to calculate the 
$\nu_e$ and $\neb$ survival probabilities. These survival probabilities
do not depend on $\delta m^2_s$ since we are 
neglecting Earth matter effects~\cite{dighe}:
\begin{eqnarray}
P_{N}(\nu_e\rightarrow\nu_e)&=&\sin^2\theta_s\cos^2\theta_x P_H
+\sin^2\theta_x(1-P_H)\,,\\
P_{N}(\neb\rightarrow\neb)&=&\cos^2\theta_s\cos^2\theta_x\,,\\
P_{I}(\nu_e\rightarrow\nu_e)&=&\sin^2\theta_s\cos^2\theta_x\,,\\
P_{I}(\neb\rightarrow\neb)&=&\cos^2\theta_s\cos^2\theta_x P_H
+\sin^2\theta_x (1-P_H)\,.
\end{eqnarray}
Here, the $N$ and $I$ subscripts denote normal and inverted hierarchy, 
respectively. We note that the factorization of the 3-neutrino 
dynamics into two 2-neutrino subsystems continues to hold with the subsystem
governed by $\delta m^2_s$ and $\theta_s$ remaining adiabatic (for the now 
well-established Large Mixing Angle solution~\cite{Eguchi:2002dm}) as in the case
of a static density profile~\cite{fogli}.

\section{Neutrino spectra}

We use the parameterization of Ref.~\cite{raffsn} for the 
primary unoscillated neutrino spectra given by
\begin{equation}
F_i(E,t)=\frac{\Phi_i(t) \beta_i^{\beta_i}}{\vev{E_i}\Gamma(\beta_i)}
\bigg(\frac{E}{\vev{E_i}}\bigg)^{\beta_i-1}
\exp\bigg(-\beta_i \frac{E}{\vev{E_i}}\bigg)\,,
\end{equation}
where $i=\nu_e$, $\neb$, $\nu_x$, and $\vev{E_i}$ ($\beta_i$) is the 
average energy (dimensionless shape parameter that quantifies the width of 
the spectrum) of species $i$. In principle, both $\vev{E_i}$ and $\beta_i$ 
can be time-dependent. Throughout, we assume $\beta_{\nu_e}=4$, 
$\beta_{\neb}=5$ and $\beta_{\nu_x}=4$~\cite{raffsn}. 
Here, $\Phi_i(t)$ is the neutrino emission
rate or luminosity of species $i$. 
The only available simulation that tracks neutrino 
emission for times long enough to enable studies of
 shock effects on neutrino oscillations
is that of the Livermore group. We adopt the luminosities of 
Ref.~\cite{totani}. 

Our goal is to demonstrate the model-independence of our new method. To this 
end, we consider models that span a wide range of predictions for the 
energy spectra. 
We conservatively consider the following parameter ranges for the initial
spectra (indicated by the 
``0'' superscript) in our analyses (with
all energies in MeV):
\begin{equation} 
14 \leq \vev{E_{\bar{\nu}_e}^0} \leq 22\,, \ \ \ 
\vev{E_{\nu_e}^0} = (0.5 - 0.9) \vev{E_{\bar{\nu}_e}^0}\,, \ \ \
\vev{E_{\nu_x}^0} = (1.0 - 1.6) \vev{E_{\bar{\nu}_e}^0}\,,
\label{energyranges}
\end{equation}
\begin{equation}
0.5 \leq \Phi_{\nu_e}^0/\Phi_{\nu_x}^0 \leq 2.0\,, \ \ \  
0.5 \leq \Phi_{\bar{\nu}_e}^0/\Phi_{\nu_x}^0 \leq 1.6\,. 
\label{lumranges}
\end{equation} 
This is in line with the hierarchy of average energies 
$\vev{E_{\nu_e}^0} < \vev{E_{\bar{\nu}_e}^0} < \vev{E_{\nu_x}^0}$, expected
from the well-known interaction strengths of neutrinos with matter; the species
which interacts less decouples earlier and has a higher temperature.

We will occasionally refer to a specific model which appears to be quite
pessimistic for determining oscillation parameters because the different 
neutrino species
have very small spectral differences. It is a model obtained
by the Garching group by accounting for all relevant
neutrino interaction effects. The initial spectra predicted by the Garching
model have~\cite{newmodel5}
\begin{equation} 
\vev{E_{\bar{\nu}_e}^0}=\vev{E_{\nu_x}^0}=15\ \rm{MeV}\,, \ \
 \vev{E_{\nu_e}^0}=12\ \rm{MeV}\,, \ \ 
\Phi_{\nu_e}^0/\Phi_{\nu_x}^0=0.5\,, \ \ 
\Phi_{\bar{\nu}_e}^0/\Phi_{\nu_x}^0=0.5\,. 
\label{Garching}
\end{equation}

\section{Simulations}

For the 1 Mton Hyper-Kamiokande detector~\cite{hyper} 
we assume an energy threshold of 
7 MeV and a fiducial volume
for SN neutrinos of 685 kton which is consistent with the fiducial to
total volume ratio expected for the UNO detector~\cite{uno}. 
We only consider events
from inverse $\beta$-decay and use the higher-order cross section of 
Ref.~\cite{beacom}. We assume that the energy resolution is that
of the Super-Kamiokande detector~\cite{Fukuda:2002pe}.

For the 100 kton liquid Ar detector~\cite{cline} with near-perfect 
efficiency, 
we consider the absorption process,
$\nu_e+ ^{40}Ar \rightarrow e^{-}+ ^{40}K^*$, whose cross section is 
calculated in Ref.~\cite{vogel}. The effective threshold is 6 MeV. 
Since the energy resolution is expected to be extremely good, we do not 
include resolution effects.

For each mass hierarchy and several values of $\theta_x$, we randomly choose
SN models defined by the parameter ranges of 
Eqs.~(\ref{energyranges},\ref{lumranges}) and
simulate $10,000$ possible 
$\nu_e$ and $\neb$ spectra at the two detectors (including statistical 
fluctuations). We fix 
$\tan^2\theta_s=0.4$~\cite{Eguchi:2002dm} and 
$|\delta m^2_a|= 0.002$ eV$^2$~\cite{superkatm}. 
We ignore Earth matter effects (thereby not
committing ourselves to specific zenith angles for the SN) which only
become comparable to shock effects at very small $\theta_x$. 

\begin{figure}[ht]
\centering\leavevmode
\includegraphics[width=6in]{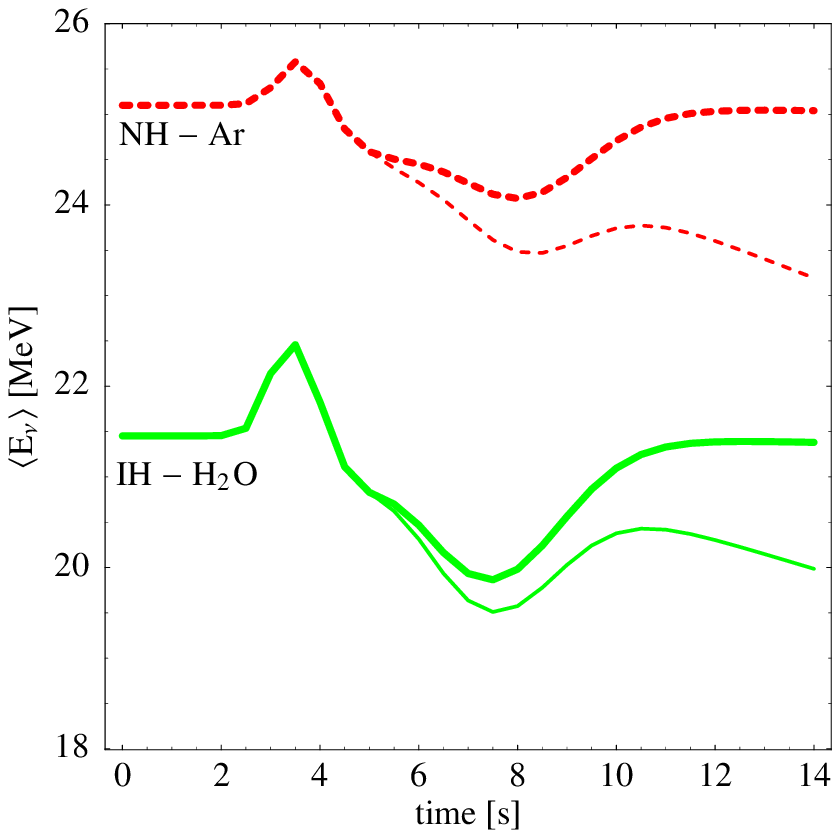}
\caption[]{The expected $\nu_e$ (solid green curves) 
and $\neb$ spectra (dashed red curves) for the normal hierarchy and inverted
hierarchy, respectively with 
$\tan^2 \theta_x=0.01$.  The thick (thin) lines correspond to 
$\vev{E_i}=\vev{E_i^0}$ ($\vev{E_i}$ that fall 
linearly 5 secs after the core-bounce).  
The curves are based on the (pessimistic) Garching model of 
Eq.(\ref{Garching}). 
Note the pronounced dips in the $\nu_e$ ($\neb$) spectra for the 
normal (inverted) hierarchy. \label{fig:spectra}}
\end{figure}

We illustrate the effect of the shock on the neutrino spectra in 
Fig.~\ref{fig:spectra}.

\section{Statistical analyses}

We consider two cases (a) $\vev{E_i}=\vev{E_i^0}$ and (b) with $\vev{E_i}$
falling linearly with time 5 secs post-bounce according to 
$\vev{E_i(t>5)}=\vev{E_i^0}(1.04167-0.008333t)$. We have chosen this 
parametrization to agree with that in Ref.~\cite{tomas}. Although 
this fall-off in energy with time 
is arbitrary, it serves to demonstrate the robustness
of our analysis to specific assumptions about $\vev{E_i(t)}$. 
After all, in a real SN, nature will make a unique selection 
for $\vev{E_i(t)}$ which is as yet unknown to us.

So long as $\tan^2 \theta_x \gg 10^{-5}$, we expect a dip in the $\nu_e$
($\neb$) spectrum if the hierarchy is normal (inverted) as in 
Fig.~\ref{fig:spectra}.
A way to establish the channel in which the dip occurs exploits the fact
that the dip is expected to occur 4 seconds after the shock forms
({\it {i.e.}} after the core rebounds). We split 
each spectrum into 2 time bins. All the events occuring in the first 4 seconds
constitute the ``early time bin'', with mean energy $\vev{E_{etb}}$ 
and the events occuring between
4 and 10 seconds constitute the ``late time bin'', with mean energy 
$E_{ltb}$. Their ratio is
\begin{equation}
R=\vev{E_{ltb}}/\vev{E_{etb}}\,.
\label{R} 
\end{equation}
In the case that $\vev{E_i}=\vev{E_i^0}$, 
for the normal hierarchy we expect $R(\nu_e,Ar)$ to be smaller than unity 
and $R(\neb,H_2O)$ to be close to unity and vice-versa for the inverted 
hierarchy. If $\tan^2 \theta_x \lsim 10^{-5}$, we expect both ratios to be
close to unity. 

For falling $\vev{E_i}$, we expect both $R(\nu_e,Ar)$ and 
$R(\neb,H_2O)$ to be slightly below unity even if 
$\tan^2 \theta_x \lsim 10^{-5}$. This is because $\vev{E_i}$ starts falling
5 seconds after bounce, which lowers $\vev{E_{ltb}}$ relative to 
$\vev{E_{etb}}$. This shock-independent suppression cannot be large since
even at $t=10$ secs, $\vev{E_i}=0.96 \vev{E_i^0}$. (Clearly the size of the
effect depends on how fast $\vev{E_i}$ falls).
Then, the hierarchy is deduced by comparing the
relative deviations of $R$ from unity. If $R(\neb,H_2O) < R(\nu_e,Ar) < 1$,
the hierarchy is inverted. If $R(\nu_e,Ar) < R(\neb,H_2O) < 1$, the hierarchy 
is normal.
 
\begin{figure}[ht]
\centering\leavevmode
\includegraphics[width=6in]{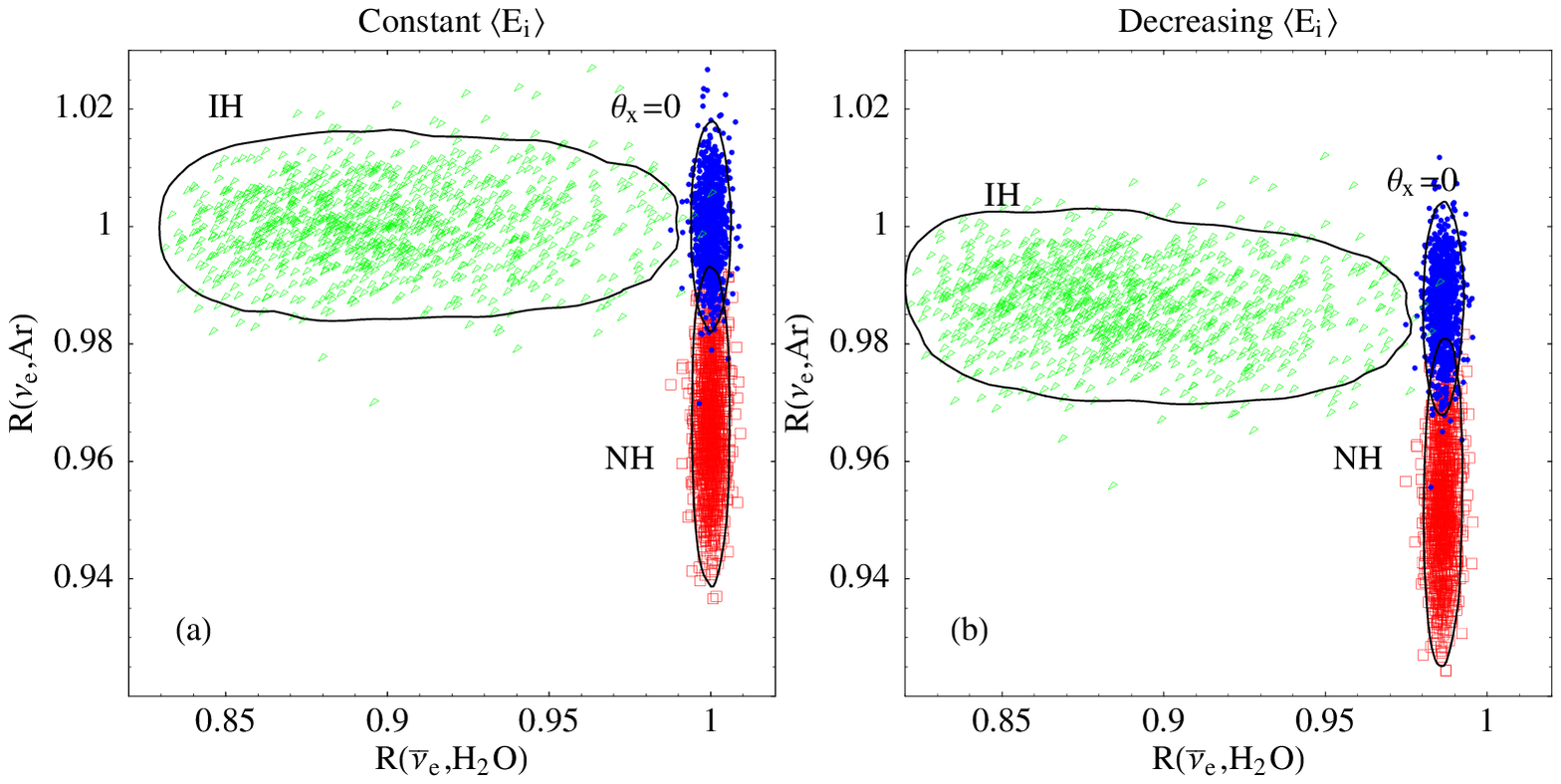}
\caption[]{(a) $R(\nu_e,Ar)$ vs. $R(\neb,H_2O)$ for SN models spanned by 
Eqs.~(\ref{energyranges},\ref{lumranges}) and $\tan^2 \theta_x=0.01$
assuming $\vev{E_i}=\vev{E_i^0}$. $R$ is the ratio of the mean energy of
the events occuring between 4 and 10 seconds to the mean energy of the events 
occuring in the first 4 seconds. The red squares
(green triangles) correspond to the 
normal (inverted) hierarchy. The blue dots are
for $\theta_x=0$. Each contour contains 95\% 
of the $10^5$ $\nu_e$ and $\neb$ 
spectra generated for the corresponding case. We have plotted only $10^3$ 
triangles, squares and dots so as to not overwhelm the figure. 
(b) is the analog of (a) but for the case in which $\vev{E_i}$ falls linearly
with time 5 secs after bounce.
The separation of the clusters is a measure of the significance with which the
hierarchy can be determined. 
\label{fig:rat}}
\end{figure}

In Fig.~\ref{fig:rat}a, 
we plot $R(\nu_e,Ar)$ vs. $R(\neb,H_2O)$ for the normal hierarchy 
(red squares) and for the inverted hierarchy (green triangles)  with 
$\tan^2 \theta_x=0.01$ assuming $\vev{E_i}=\vev{E_i^0}$. 
The blue dots are for $\theta_x=0$.  
There is little 
overlap between the clusters, and the inverted
hierarchy can be established at more than 3$\sigma$. 
As can be seen in Fig.~\ref{fig:rat}b, 
the case in which $\vev{E_i}$ fall linearly also yields
little overlap between the clusters, although the clusters are now 
systematically displaced to lower values of $R(\nu_e,Ar)$ and $R(\neb,H_2O)$.

\begin{figure}[ht]
\centering\leavevmode
\mbox{\includegraphics[width=6in]{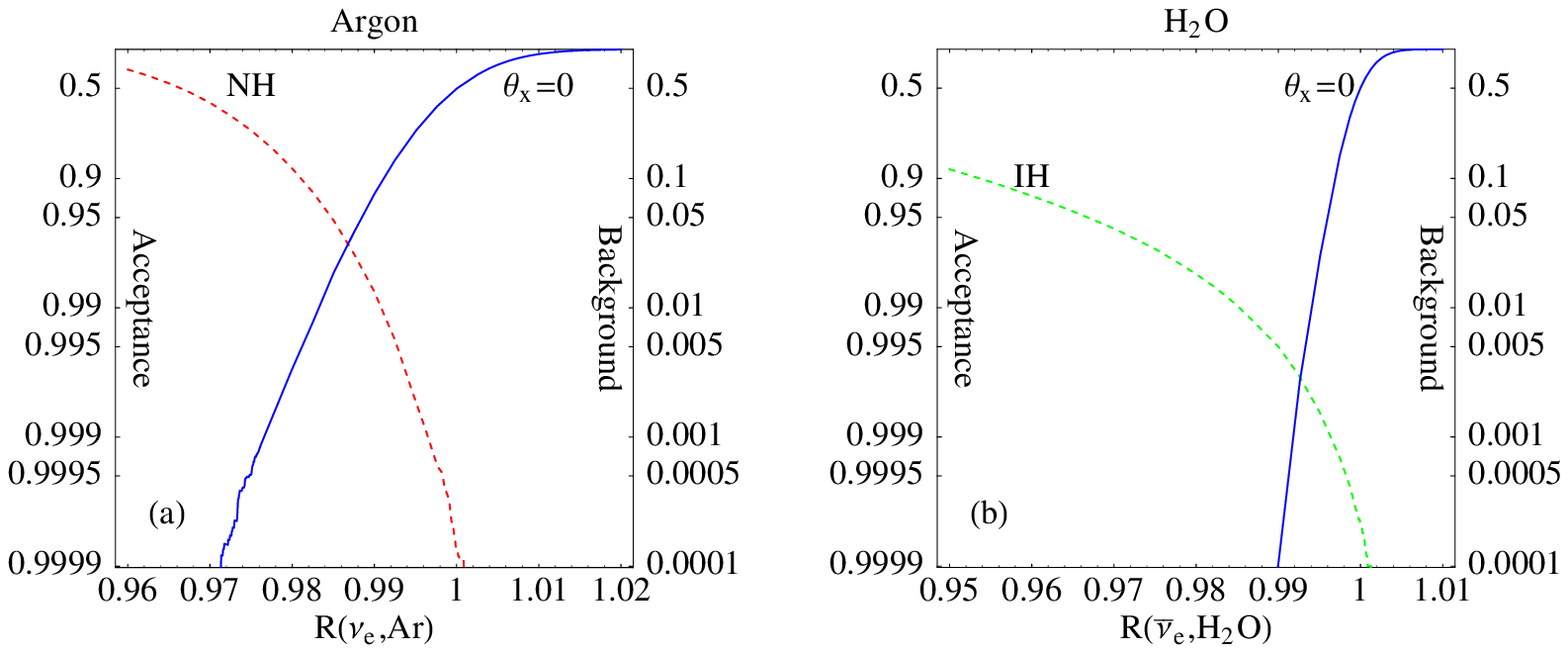}}
\caption[]{(a) The dashed red
curve shows the fraction of 
spectra corresponding to the normal hierarchy and 
$\tan^2 \theta_x=0.01$ with $R(\nu_e,Ar)$ below a given value 
(use the y-scale on the left).
The solid blue curve shows the number of spectra 
with $\theta_x=0$ and $R(\nu_e,Ar)$
below a given value 
divided by the number for the normal hierarchy and $\tan^2 \theta_x=0.01$
with $R(\nu_e,Ar)$ below the same value (use the y-scale on the right). 
(b) is 
the analog of (a) but for the inverted hierarchy. Spectra with 
$\vev{E_i}=\vev{E_i^0}$ were assumed.  
See Eq.~(\ref{R}) for the definition of $R$.
\label{fig:sig}}
\end{figure}

We now focus on the $\vev{E_i}=\vev{E_i^0}$ case.
It is no surprise that 
it is more difficult to identify
the normal hierarchy considering the lower statistics in the Ar detector.
However, it is possible to establish the normal hierarchy 
reasonably well. 
In Fig.~\ref{fig:sig}a the dashed red
curve shows the fraction of spectra corresponding to the normal hierarchy and 
$\tan^2 \theta_x=0.01$ with $R(\nu_e,Ar)$ below a given value. For example,
88\% (98.7\%) of the spectra corresponding to the normal hierarchy and 
$\tan^2 \theta_x=0.01$ have $R(\nu_e,Ar)<0.98$ (0.99).
 The solid blue 
curve shows the number of spectra with $\theta_x=0$ and $R(\nu_e,Ar)$
below a given value 
divided by the number for the normal hierarchy and $\tan^2 \theta_x=0.01$
with $R(\nu_e,Ar)$ below the same value. Thus from Fig.~\ref{fig:sig}a,
 the fraction of spectra with 
$\theta_x=0$ that mimic the spectra involving shocks, 
and have $R(\nu_e,Ar)<0.98$ (0.99), is 0.33\% (7.6\%).
Said differently, 88\% (98.7\%) of the spectra indicate the normal 
hierarchy correctly with 
a probability of 99.67\% (92.4\%).  

We emphasize that these results have included the Garching model 
of Eq.~(\ref{Garching}).  
As expected, the results of
a similar analysis of $1,000$ spectra 
simulated from the Garching model alone are somewhat worse. 
Specifically, 
we find that  87\% (99\%) of the 
spectra simulated with the normal hierarchy and $\tan^2\theta_x=0.01$ have
$R(\nu_e,Ar)<0.98$ (0.99),
and identify the normal hierarchy with a confidence of 99\% (87\%).
It is remarkable that even for initial spectra with tiny differences, such 
as in the Garching model, it
is possible to establish the normal hierarchy.

Figure~\ref{fig:sig}b is similar to~\ref{fig:sig}a, 
but for the inverted hierarchy.
Clearly, a determination of the inverted hierarchy is easier. 
We see that  99.5\% of the spectra simulated with an inverted
hierarchy and $\tan^2\theta_x=0.01$ 
have $R(\neb,H_2O)<0.99$, 
and indicate the inverted hierarchy at the 99.99\% C.~L. 

 The case in which
$\vev{E_i}$ fall linearly with time 5 seconds after bounce 
gives similar results as those 
for the constant $\vev{E_i}$ case.

We have also investigated if the hierarchy can be determined for 
$\tan^2 \theta_x=0.001$. It is still possible to establish the the inverted
hierarchy with high-confidence. For the normal hierarchy we find that 
while shock effects remain visible, it is no longer possible to establish the
normal hierarchy with high-significance. 

\section{Conclusions}

Our analysis has established that detections of shock modulations 
in either the $\nu_e$ spectrum in a kton-class liquid Ar detector or 
in the $\neb$ spectrum in a Mton-class water Cherenkov detector
will provide definitive evidence
for the true neutrino mass hierarchy and that 
$\tan^2 \theta_x \gg 10^{-5}$. 
We emphasize that the modulations should be observed 
in one and only one of the two channels. 

A similar analysis can be performed supposing that a reverse shock also
develops in the SN envelope~\cite{moresn}. 
Then, one expects to see a double-dip
signature in either the $\nu_e$ or $\neb$ spectrum if $\theta_x$ is not 
too small~\cite{tomas}. In addition to probing oscillation parameters,
a presence of a double 
dip will provide confirmation that SN simulations correctly depict 
shock propagation.   

We have supposed that both a large water Cherenkov and 
large liquid argon detector are available. 
A possible alternative to a 
large water Cherenkov detector is the IceCube detector since
good energy resolution is not a requirement of our analysis.
Then $R(\neb,Icecube)$ can be compared with $R(\nu_e,Ar)$~\cite{moresn}.
 
An aspect that bears further consideration is whether our analysis 
is robust under
variations of the shock density profile. We are currently investigating this
issue in Ref.~\cite{moresn}.

\section{Acknowledgments}

We thank H.-T.~Janka, G.~Raffelt and R.~Tomas for useful discussions and
for providing us with data on profiles and luminosities.
This research was supported by the U.S. Department of Energy under
Grant No.~DE-FG02-95ER40896, by the NSF under Grant No.~EPS-0236913, by
the State of Kansas through Kansas Technology Enterprise Corporation and by the
Wisconsin Alumni Research Foundation.

\end{document}